\documentclass[iop,dvipdfmx]{emulateapj}

\usepackage{bm}
\usepackage{float}
\usepackage{amsmath}
\usepackage{mathrsfs}

\usepackage{upgreek}
\usepackage{graphicx}
\usepackage{color}

\usepackage{threeparttable}

\newcommand{\msunyr}{M_{\odot}\:{\rm yr}^{-1}}
\newcommand{\mdot}{\dot{m}}
\newcommand{\Mdot}{\dot{M}}
\newcommand{\msun}{M_{\odot}}

\newcommand{\zsun}{Z_{\odot}}

\newcommand{\Sigcl}{\Sigma_{\rm cl}}
\newcommand{\gcm}{{\rm g\:cm^{-2}}}

\newcommand{\kap}{\kappa}
\newcommand{\sfe}{\bar{\varepsilon}_{\rm *f}}
\newcommand{\N}{\mathscr{N}}

\defcitealias{KT17}{Paper I}

\renewcommand{\thefootnote}{\fnsymbol{footnote}}

\shorttitle{Feedback in Low-Metallicity Massive Star Formation}
\shortauthors{Tanaka, Tan, Zhang, \& Hosokawa}

\begin{document}

\title{The Impact of Feedback in Massive Star Formation. II.\\
Lower Star Formation Efficiency at Lower Metallicity}

\author{Kei E. I. Tanaka}
\affil{Department of Earth and Space Science, Osaka University, Toyonaka, Osaka 560-0043, Japan; ktanaka@astro-osaka.jp}
\affil{Chile Observatory, National Astronomical Observatory of Japan, Mitaka, Tokyo 181-8588, Japan}
\affil{Department of Astronomy, University of Florida, Gainesville, FL 32611, USA}
\author{Jonathan C. Tan}
\affil{Department of Space, Earth and Environment, Chalmers University of Technology, SE-412 96 Gothenburg, Sweden}
\affil{Department of Astronomy, University of Virginia, Charlottesville, VA 22904, USA}
\author{Yichen Zhang}
\affil{Star and Planet Formation Laboratory, RIKEN Cluster for Pioneering Research, Wako, Saitama 351-0198, Japan}
\and
\author{Takashi Hosokawa}
\affil{Department of Physics, Kyoto University, Sakyo, Kyoto 6060-8502, Japan}

\begin{abstract}
We conduct a theoretical study of the formation of massive stars over
a wide range of metallicities from $10^{-5}$ to $1\zsun$ and evaluate
the star formation efficiencies (SFEs) from prestellar cloud cores
taking into account multiple feedback processes.  Unlike for simple
spherical accretion, in the case of disk accretion feedback processes
do not set upper limits on stellar masses.
At solar metallicity, launching of magneto-centrifugally-driven outflows is the
dominant feedback process to set SFEs, while radiation pressure, which
has been regarded to be pivotal, has only minor contribution even in
the formation of over-$100\:\msun$ stars.  Photoevaporation becomes
significant in over-$20\:\msun$ star formation at low metallicities of
$\la10^{-2}\:\zsun$, where dust absorption of ionizing photons is
inefficient.  We conclude that if initial prestellar core properties
are similar, then massive stars are rarer in extremely metal-poor
environments of $10^{-5}$--$10^{-3}\:\zsun$.  Our results give new
insight into the high-mass end of the initial mass function and its
potential variation with galactic and cosmological environments.
\end{abstract}

\keywords{stars: massive, formation, evolution, mass function, outflows}

\section{Introduction} \label{sec_intro}

Massive stars are the main sources of UV radiation, turbulent energy,
and heavy elements.  Massive close-binaries are the progenitors of
merging black holes which have been detected by their gravitational
wave emission.  Even though they play crucial roles across a wide
range of astrophysics, the formation of massive stars is still not
fully comprehended.  Especially, it is important to understand how the
massive star formation process depends on galactic environmental
conditions, since this shapes the high-mass end of the initial mass
function (IMF) and affects how the IMF may vary through cosmic
history.

To address this topic, we have developed a model of feedback during
massive star formation relevant for solar metallicity conditions,
especially star formation efficiencies from prestellar gas cores
\citep[][hereafter Paper I]{KT17}.  Here we apply this model to a
wide range of metallicities that are expected to be relevant to
galactic environments across most of the evolution of the universe.

\begin{figure*}
\begin{center}
\includegraphics[width=180mm]{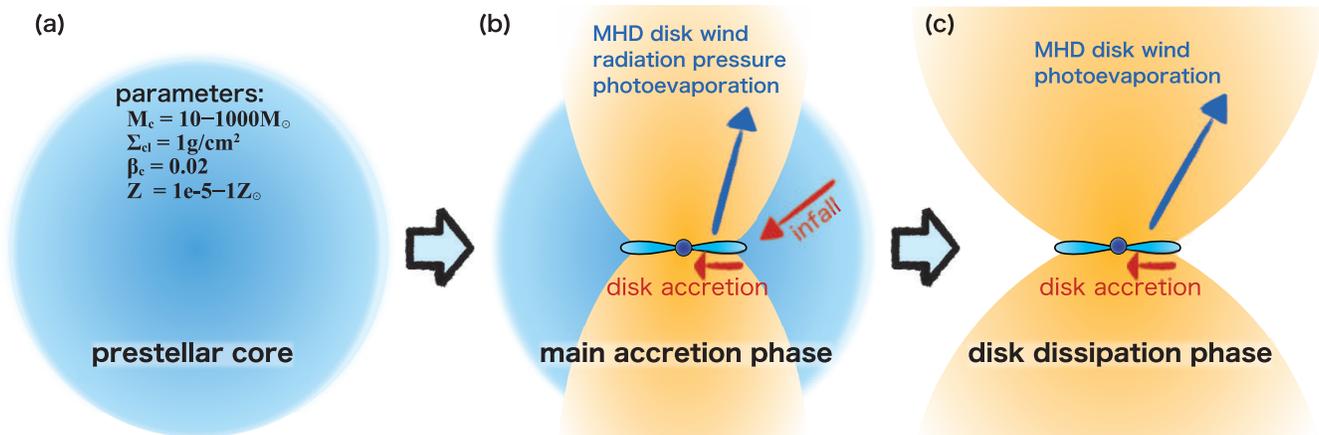}
\end{center}
\caption{
Overview of the evolutionary stages of massive star formation in our
model, which is based on the Core Accretion paradigm. {\it (a):} The
initial prestellar cloud core is spherical and close to virial
equilibrium. The structure is characterized by three main parameters:
core mass, $M_c$; mass surface density of ambient clump, $\Sigcl$; and
the ratio of core's initial rotational to gravitational energy
$\beta_c$ \citep{mck03}.
Here we assume metallicity, $Z$, may alter
feedback effects, but not core structure and accretion properties.
{\it (b):} In the main accretion phase, the infalling envelope
accretes onto the central protostar through the disk. The outflow
cavity is opened up by the momentum of an MHD disk wind, with later
contributions from radiation pressure, leading to reduction of the
solid angle of the region that is able to infall. Additionally, mass
loss by the MHD disk wind and photoevaporation reduces the accretion
rate onto the star.  {\it (c):} When infall from the envelope is
finished, the disk starts to dissipate by mass accretion onto the star
and mass loss caused by the MHD disk wind and photoevaporation. The
stellar birth mass, in the approximate limit of formation of a single
dominant star, is set when the remnant disk has finally dissipated.  }
\label{fig_schematics}
\end{figure*}

Radiative feedback has been considered to be critical in setting the
mass of massive stars at their birth.  Especially, radiation pressure
acting on dust grains has been modelled to be a potential barrier to
the formation of present-day massive stars.  For the idealized case of
spherical accretion, radiation pressure acting on the dusty envelope
exceeds gravitational attraction when the stellar mass reaches about
$20\:\msun$ preventing further mass accretion \citep{lar71,wol87}.
The existence of more massive stars indicates that a non-spherical
accretion geometry, i.e., involving accretion disks, is important.
Subsequent studies via (semi-)analytic models
\citep{nak89,jij96,tan11} and numerical simulations \citep{yor02,
  kru09, kui10, ros16} showed that the gas behind the disk, which is
expected to be optically thick, is shielded from radiation pressure
and thus accretion can continue to high masses, potentially $\gtrsim
100\:M_\odot$, depending on the initial condition of the core.
Radiation predominantly escapes via lower density cavities above and
below the disk.  Such low-density cavities may be opened by the
stellar radiation itself \citep{kui10,kui11}, including via radiative
Rayleigh-Taylor instabilities \citep{kru09,ros16} or, more likely,
by magneto-centifugally-driven outflows
\citep{kru05,kui15,kui16,mat17}.

As a result of the above studies, radiation pressure is no longer
regarded as a feedback mechanism that is catastrophic for massive star
formation.  Still, there is a more general remaining question about
the quantitative effects of feedback mechanisms in setting star
formation efficiencies from gas cores, potentially shaping the stellar
IMF and its variation with metallicity.

In the formation of primordial (Pop III) stars, i.e., the limit of
zero metallicity, radiation pressure is not expected to be significant
because there are no dust grains.  Instead of radiation pressure,
photoevaporation is thought to be a critical feedback process for
setting the mass of Pop III stars.  As a massive primordial protostar
approaches the Zero-Age Main-Sequence (ZAMS), it starts to emit vast
amounts of Lyman continuum photons, i.e., with $>13.6\rm\:eV$ that
would ionize infalling and accreting material.  The thermal pressure
of such ionized gas with $\gtrsim10^4\rm\:K$ drives a photoevaporative
flow \citep{hol94}, which staunches mass accretion at stellar masses
of $\sim50$--$100\:\msun$ \citep{mck08,hos11,KT13,KT17}.
Photoevaporation may also be important in the formation of massive
stars in non-zero metallicity environments.  Recently, \citet{nak18}
performed radiative hydrodynamical simulations showing the metallicity
dependence of the photoevaporation rate.  However, their focus is on
the dissipation of protoplanetary disks around low-mass protostars
with $0.5\:\msun$.  Although there are some similarities, their model
is not applicable to our study because the luminosity and the spectrum
are quite different between low- and high-mass stars.  It is still
uncertain how photoevaporation feedback during massive star formation
depends on metallicity.

Non-radiative feedback, namely magneto-centifugally-driven outflows,
may also be important.  In the mass range lower than $10\:\msun$
and in local Milky Way environments, the observed core mass function
(CMF) is reported to be similar in shape to the stellar IMF, but with
a shift to higher masses by a factor of a few
\citep[e.g.,][]{and10,kon10}. One promising explanation for this is
that SFEs from prestellar cores may be $\sim0.4$ for both low- and
intermediate-mass star formation. Theoretical and numerical studies of
low-mass star formation proposed this SFE value is set by outflow
feedback that is driven by the momentum of a magnetohydrodynamic (MHD)
disk wind \citep{mat00,mac12,zha15,off17}.  In the formation of
massive stars, on the other hand, theoretical studies have paid most
attention to radiative feedback because of their enormous
luminosities.  However, observations suggest that the structures of
the outflows around low- and high-mass protostars are similar
\citep[e.g.,][]{qiu09,zha13a,deb17,hir17,mcl18}.  The models of
\citet{zha13,zha14,zha18} have considered the formation of
massive stars from cores with the only feedback effect included being
that due to MHD outflows.  Scaling up the assumptions of the model of
\citet{mat00}, they find similar SFEs from massive cores of $\sim0.5$.
\citet{mat17} recently performed MHD simulations of the collapse
of massive magnetized cloud cores, ignoring radiative feedback.  They
showed that an MHD outflow is launched in a similar way to the case
of low-mass star formation, but is more powerful due to the higher
accretion rate and deeper gravitational potential. Hence MHD outflow
feedback is expected to also play an important role in massive star
formation.

In reality, massive stars are formed under the influence of all of
these feedback processes. Paper I studied the impact of multiple
feedback processes in massive star formation using semi-analytic
methods and found that MHD disk wind feedback is more important
compared to radiative feedback.
In this sense and under the
assumptions of the modeling via Core Accretion \citep{mck03}
the formation of massive stars is similar to those of low-mass stars.
Recently \citet{kui18} also studied the combination of multiple feedback processes
(disk winds, radiation pressure and photoionization)
by radiative-hydrodynamical simulations, which well agreed with our semi-analytic work
of \citetalias{KT17}. 
However, in this paper
we focused mainly on present-day massive star formation assuming solar
metallicity. In this paper, to investigate how the formation processes
of massive stars change with galactic environment and over cosmic
history, we extend our model to lower metallicities and evaluate the
impact of feedback and SFEs from given prestellar cloud cores.  We
note that, with a similar conceptual framework, \citet{hos09b} and
\citet*{fuk18} have studied radiative feedback and estimated the
maximum stellar mass as a function of metallicity.  However, they
assumed spherical accretion geometry and ignored the MHD disk wind.
As described above, disk accretion is a key factor to circumvent the
radiation pressure barrier, and MHD disk winds likely have
considerable contributions in massive star formation.  We adopt an
axisymmetric model allowing treatment of these processes,
which we will see leads to completely different outcomes from the idealized
spherical calculations.

This paper is organized as follows. In \S\ref{sec_method} we review
the basics of our model and introduce the updates needed to treat the
effects of feedback processes at a range of metallicities.  Then, in
\S\ref{sec_results} we present our results: we show the metallicity
dependencies of the feedback processes and the SFE.  In
\S\ref{sec_discussion} we discuss the potential implications for IMF
variation based on the obtained SFE model.  We conclude in
\S\ref{sec_conclusions}.

%%%%%%%%%% Section 2 %%%%%%%%%%

\section{Methods}\label{sec_method}

We calculate the accretion history onto massive protostars including
effects of several feedback processes. Figure \ref{fig_schematics}
presents a schematic overview of our model. In \citetalias{KT17}, we
focused on massive star formation at solar metallicity.  Our model is
developed under the paradigm of the Turbulent Core Model
\citep{mck03}, which is a Core Accretion model scaled-up from those
developed for low-mass star formation.
We now extend this model in the metallicity range from $10^{-5}\zsun$
to $1\:\zsun$.  The main update from \citetalias{KT17} is taking into
account how metallicity influences protostellar evolution and
radiative feedback.  Here we review the basics components of the model
and these updates.

\subsection{Prestellar Cloud Cores}\label{sec_core}

Our model assumes single star formation from a collapsing prestellar
cloud core (Figure \ref{fig_schematics}a).  The initial core is
spherical and close to virial equilibrium, being supported by
non-thermal pressure components, i.e., turbulence and magnetic fields.
The core properties in this model are characterized by three
fundamental parameters: core mass, $M_c$; mass surface density of the
ambient clump, $\Sigma_{\rm cl}$; and ratio of the core's initial
rotational to gravitational energies, $\beta_c$.  The size of the core
is set as
$R_c=0.057(M_c/60\:\msun)^{1/2}(\Sigcl/1\:\gcm)^{-1/2}{\rm\:pc}$ under
the assumption of pressure equilibrium of the core surface with the
ambient clump medium, and with the normalization factor for a
power-law internal core density profile $\rho\propto r^{-1.5}$
\citep[observations of dense cores in Infrared Dark Clouds indicate
a density power law index of $\simeq1.3$--$1.6$,][]{but12,but14}.
In this study, we investigate core masses in the range of
$M_c=10$--$1000\msun$, fixing the clump mass surface density and the
rotational parameter at fiducial values, i.e., $\Sigcl=1\:\gcm$
\citep{plu97,mck03,tan14} and $\beta_c=0.02$ \citep{goo93,li12,pal13}.
We note that we assume that metallicity does not alter the properties
of prestellar cores in order to clarify the influence of metallicity
on the feedback processes.
We will discuss how core properties may vary at various metallicities in \S\ref{sec_discussion}.

\subsection{The Main Accretion Phase}

The prestellar core undergoes gravitational collapse forming a
protostar at its center.  Material accretes onto the star through the
disk under the influence of multiple feedback process (Figure
\ref{fig_schematics}b).  In this work, we consider feedback due to MHD
disk winds, radiation pressure and photoevaporation, as explained
later in this section.  In \citetalias{KT17}, we also considered mass
loss by stellar winds using a wind model for hot stars with
$30,000$--$50,000\rm\:K$ \citep{vin11} and found that stellar wind
rates are several orders of magnitude smaller than those by other
processes at the solar metallicity.  The metal-line-driven stellar
winds are expected to be even weaker at lower metallicities.
Recently, \citet{vin18} showed that, in the case of inflated very
massive stars with a relatively low effective temperatures of
$\sim15,000\rm\:K$, stellar wind mass loss rates can reach as high as
$10^{-3}\:\msunyr$ at stellar masses $>800 (Z/\zsun)^{-0.35} \msun$.
However, in our model setup, stars above $100\:\msun$ always have a
high effective temperature of $\ga30,000\rm\:K$.  Therefore, in this
study, the mass loss from stellar winds is not considered.

\subsubsection{Infall, Disks and Protostars}\label{sec_infall+}

The infall of the core is described by the self-similar solution
\citep{mcl97,mck03}, which gives the infall rate onto the
protostar-disk system in the limit of {\it no feedback}:
\begin{eqnarray}
\Mdot_{\rm *d}(t)&=&9.2\times10^{-4}
\left( \frac{M_{\rm *d}}{M_c} \right)^{0.5} \nonumber\\
&\times& \left( \frac{M_c}{60\:\msun} \right)^{3/4}
\left( \frac{\Sigcl}{1\:\gcm} \right)^{3/4}\msunyr, \label{eq_infall}
\end{eqnarray}
where $M_{\rm *d}=\int \Mdot_{\rm *d}dt$ is the collapsed mass, which is
the total mass of the protostar and disk in the limit of no feedback.
The obtained infall rate is orders of magnitude higher compared to the
typical accretion rates of $\sim10^{-6}\msunyr$ that are considered to
be characteristic of low-mass star formation. However, the actual accretion
rate onto the protostar is reduced from the value in Equation
(\ref{eq_infall}) due to feedback processes (\S\ref{sec_accretion}).

A disk is formed around the protostar because the initial core is
rotating. Assuming angular momentum conservation of infalling gas from
the sonic point, where the infall velocity reaches the sound velocity,
the disk radius is given by
\begin{eqnarray}
r_d(t)&=&156
\left( \frac{\beta_c}{0.02} \right)
\left( \frac{M_{\rm *d}}{m_{\rm *d}} \right)
\left( \frac{M_{\rm *d}}{M_c} \right)^{2/3} \nonumber\\
&\times&
\left( \frac{M_c}{60\:\msun} \right)^{1/2}
\left( \frac{\Sigcl}{1\:\gcm} \right)^{1/2}
{\rm AU}
\end{eqnarray}
\citep[see \S2.1 of][]{zha14}. The disk is massive and
self-gravitating due to high supply rate from the envelope.  The
angular momentum in the disk is efficiently transported by torques in
such a massive disk, keeping the mass ratio of the disk to protostar
approximately constant at $\sim1/3$ \citep[e.g.,][]{kra08}.
Therefore, this ratio is fixed at $f_{d}=1/3$ during the main
accretion phase following the assumption adopted in our previous
series of papers \citep[][\citetalias{KT17}]{zha11,zha13,zha14,zha18}.

To evaluate the strength of feedback, the properties of the protostar,
such as luminosity, radius and effective temperature, along with their
evolution, are important.  Therefore, we calculate the protostellar
evolution self-consistently given the accretion rate using the model
of \citet{hos09} and \citet{hos10}, which solves the basic stellar
structure equations, i.e., continuity, hydrostatic equilibrium, energy
conservation, and transport \citep{sta80,pal91}.  This model has also
performed successfully at all metallicities in the range of
$Z=0$--$1\zsun$ \citep{hos09b,fuk18}.  The stellar boundary condition
is adopted from two types depending on the accretion geometry, i.e.,
spherical or disk accretion.  In the earliest stages, the expected
disk radius, $r_d$, is smaller than the stellar radius $r_*$ and the
accretion flow is quasi-spherical.  In this case, a shock front is
produced at the stellar surface and a fraction of released flow energy
is advected into the interior, which is referred to as the {\it hot}
shock boundary.  On the other hand, if $r_d>r_*$, gas accretes onto
the stellar surface through a geometrically thin disk. Then much of
the energy is radiated away before the material settles onto the star.
Thus, in this case, the {\it cold} photospheric boundary condition is
adopted, in which the specific entropy carried into the star is
assumed to be the same as the gas at the stellar photosphere.

\subsubsection{MHD Disk Wind and Radiation Pressure}\label{sec_outflow}

A bipolar outflow sweeps up part of the core reducing the amount of
gas that can accrete onto the star.  We evaluate the opening angle of
the outflow cavity $\theta_{\rm esc}$ considering the total momenta of
the MHD disk wind and radiation pressure.  \citet{mat00} developed a
basic model of outflows driven by the momentum of an MHD disk wind,
applied in the context of low-mass star formation.
\citet{zha13,zha14} applied this model to the case of high-mass star
formation, finding MHD outflow feedback creates outflow cavities that
open-up during the course of star formation and set formation
efficiencies from the core of $\sim50\%$.  In \citetalias{KT17}, we
introduced the contribution of radiation pressure to this
momentum-driven outflow model.  The outflow cavity extends to a
certain angle if the outflow momentum is strong enough to accelerate
the core material to its escape speed in that direction: the following
equation is satisfied at the polar angle of $\theta=\theta_{\rm
  esc}(t)$
\begin{eqnarray}
c_g \frac{dM_c}{d\Omega} {\rm v}_{\rm esc}=\frac{dp_{\rm dw}(t)}{d\Omega} + \frac{dp_{\rm rad}(t)}{d\Omega},
\end{eqnarray}
where $p_{\rm dw}$ and $p_{\rm rad}$ are the momenta of the MHD disk
wind and radiation pressure, respectively, $\Omega$ is the solid
angle, ${\rm v}_{\rm esc}=\sqrt{2GM_c/R_c}$ is the escape speed from
the core.  The correction factor $c_g$ was introduced by \citet{mat00}
to account for the effect of gravity and the propagation of the
shocked shell, which is evaluated as $2.63$ for our core setup.  As
the momenta from the MHD disk wind and the radiation pressure keep
accumulating, the opening angle of the outflow increases with time
until it reaches the maximum angle that is limited by the disk aspect
ratio, i.e., $\theta_{\rm esc, max}=\tan^{-1}(H/r)$, where $H$ is the
disk scale height. Infall can always continue from the equatorial
region in the disk shadow because shielding by the inner disk region
is efficient at overcoming radiation pressure \citep{tan11,kui12}.
The inner disk aspect ratio is evaluated at the radius of $r=10r_*$
following \citet{mck08}.  The typical value of the aspect ratio is
$\sim0.1$, corresponding to a maximum opening angle of $\sim84\degr$.

\begin{figure*}
\begin{center}
\includegraphics[width=180mm]{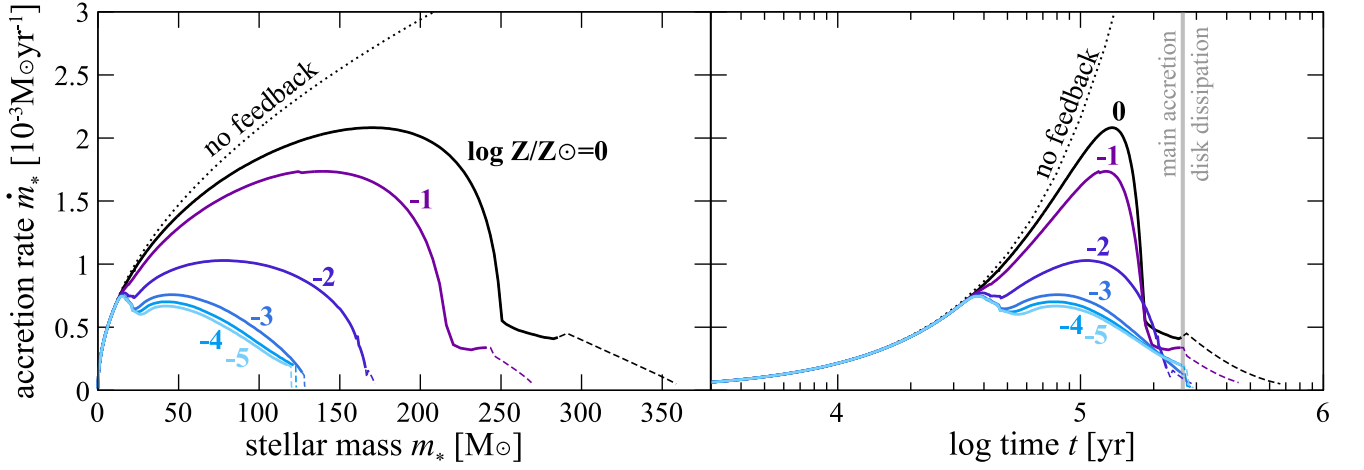}
\end{center}
\caption{
Accretion histories as functions of protostellar mass, $m_*$, ({\it
  left}) and time, $t$, ({\it right}) for stars forming from cores
with initial masses of $M_c=1000\:\msun$ and embedded in clump
environments with $\Sigma_{\rm cl}=1\:{\rm g\:cm}^{-2}$.  Results
for metallicities $\log Z/\zsun=-5,\:-4,\:-3,\:-2,\:-1,$
and $,\:0$ are shown as labelled.  In each line, the solid part
represents the main accretion phase and the dashed part is the disk
dissipation phase (the gray vertical line in the right panel indicates
the transition time).  The black dotted lines show the no feedback case for
reference.  The accretion rate is lower at lower metallicity due to
stronger total feedback.}
\label{fig_acc}
\end{figure*}

\begin{figure*}
\begin{center}
\includegraphics[width=180mm]{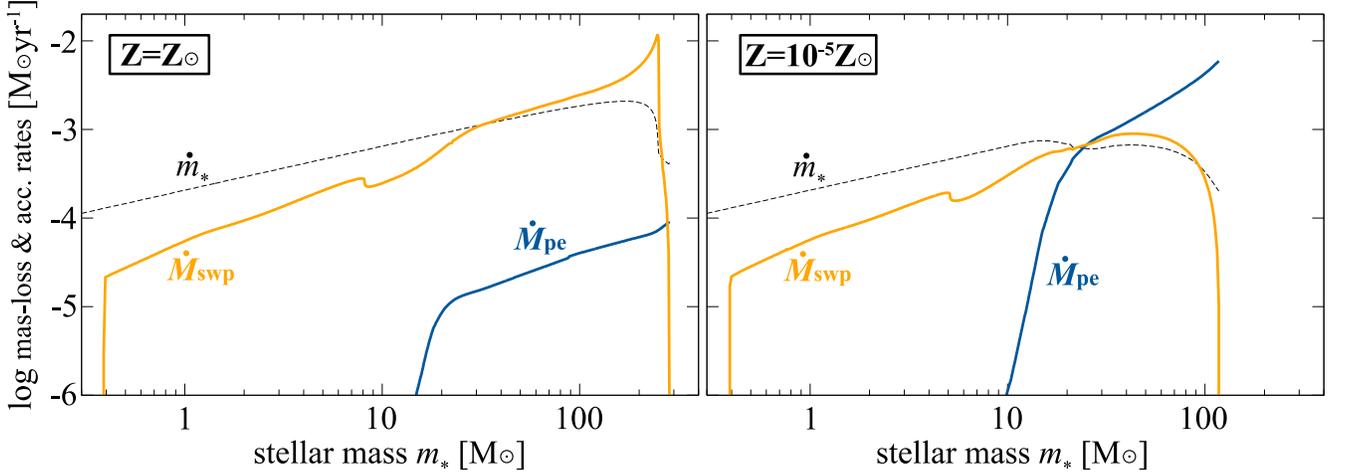}
\end{center}
\caption{
Mass-loss rates by outflow sweeping (orange solid lines) and by
photoevaporation (blue solid lines)
during the main-accretion phase at $Z=\zsun$ ({\it left panel})
and at $10^{-5}\:\zsun$ ({\it right panel}) from the same initial core
with $M_c=1000\:\msun$ and $\Sigma_{\rm cl}=1\:{\rm g\:cm}^{-2}$.  The
black dashed lines represent the accretion rates.  Outflow sweeping is
the dominant feedback at $\zsun$, while photoevaporation becomes more
significant at $10^{-5}\:\zsun$.}
\label{fig_mass_loss}
\end{figure*}

Disk winds driven magneto-centrifugally \citep{bla82} are adopted in
our study.  The mass loading fraction of the wind relative to the
accretion rate onto the star is assumed to be $f_{\rm dw}=0.1$ as a
typical value of disk winds \citep{kon00}.  Note that, due to the
trapping by the core, the actual mass-loss rate in the disk wind is
smaller than $f_{\rm dw}\mdot_*$, which is the value in the limit of a
fully opened outflow cavity.  Taking this into account, we
describe the disk wind mass-loss rate as
\begin{eqnarray}
\mdot_{\rm dw}=f_{\rm dw,esc}f_{\rm dw}\mdot_*, \label{eq_mdotdw}
\end{eqnarray}
where $f_{\rm dw,esc}$ is the fraction of the mass of the wind that
can escape from the outflow cavity, which is evaluated based on the
the mass flow in the direction $\theta\le\theta_{\rm esc}$
\citep{zha14}.  Then, the disk wind momentum $p_{\rm dw}$ is evaluated
by integrating the momentum rate of the disk wind,
\begin{eqnarray}
\dot{p}_{\rm dw}(t) = \phi_{\rm dw}\mdot_{*}{\rm v}_{\rm K*},  \label{eq_pdotdw}
\end{eqnarray}
where $\mdot_*$ is the accretion rate onto the star and ${\rm
  v}_{\rm K*}$ is the Keplerian speed at the stellar radius. The
factor of $\phi_{\rm dw}$ is introduced to measure the disk wind
momentum to $\mdot_*{\rm v}_{\rm K*}$ \citep{tan02}.  For our disk
wind model, the value of $\phi_{\rm dw}$ is about $0.15$--$0.3$.  We
note that the momentum of the MHD disk wind obtained from our analytic
model agrees well with the recent MHD simulation of massive star
formation by \citet{mat17}.  The angular distribution of the momentum
of MHD disk wind is described as \citep{shu95,ost97,mat99}
\begin{eqnarray}
P(\mu)=\frac{dp_{\rm dw}}{d\Omega} \frac{4\pi}{p_{\rm dw}}=\frac{1}{\ln \left( 2/\theta_0 \right) \left(1+\theta_0^2 - \mu^2 \right)},
\end{eqnarray}
where $\theta_0$ is a small angle that is estimated to be $0.01$ and
$\mu=\cos\theta$ (note that $\int_0^1 Pd\mu=1$).  This angular
distribution of $P(\mu)$ encapsulates the collimated nature of MHD
disk winds.  \citet{hig18} followed evolution of magnetic fields in
collapsing star-forming clouds using non-ideal simulations before the
main accretion phase starts.  However, metallicity dependences of MHD
disk winds during the accretion phase are still uncertain.  Therefore,
for simplicity, the MHD disk wind is assumed to be independent of the
metallicity in this study.  For more details about the disk wind
momentum, see also \S2.3 of \citet{zha14} and \S2.2.1 of
\citetalias{KT17}.

The momentum from radiation pressure, $p_{\rm rad}$, is obtained by the
integral of the radiation pressure momentum injection rate,
\begin{eqnarray}
\dot{p}_{\rm rad}&=&f_{\rm trap} \frac{L_{\rm *acc}}{c},
\end{eqnarray}
where $f_{\rm trap}$ is a trapping factor accounting for the optical
depth to the stellar radiation and the dust re-emission.  In the
spherical limit at solar metallicity, the infalling envelope is
optically thick not only for the direct stellar radiation but also to
the infrared radiation re-emitted from dust grains.  Then, the
trapping factor is larger than unity, i.e., $f_{\rm trap}\simeq
\uptau_{\rm IR}\gg1$ \citep{tho05}, boosting the contribution of
radiation pressure feedback.
However, in non-spherical accretion,
this radiation pressure by dust re-emission is reduced significantly
by the pre-existing MHD outflow cavity \citep{kru05,kui15,kui16}.
Therefore, in our models, the effect of dust re-emission is ignored and only
direct stellar radiation absorbed by dust grains is considered.
In \citetalias{KT17}, we assumed the envelope is optically thick to the
stellar radiation and $f_{\rm trap}=1$ because we were interested in
the solar metallicity case.  To treat the low metallicity case properly, we
modify $f_{\rm trap}$ as
\begin{eqnarray}
f_{\rm trap} &=& 1-\exp\left({-\uptau_{\rm env}}\right), \label{eq_frap} \\
\uptau_{\rm env}&=&\kap_{\rm *acc}\Sigma_{\rm env}= \kap_{\rm *acc}\int_{r_{\rm sub}}^{R_{\rm c}} \rho_{\rm env} dr,
\end{eqnarray}
where $\uptau_{\rm env}$ is the optical depth of the infalling
envelope to the direct stellar radiation, $\kap_{\rm *acc}$ is the
Planck mean opacity at the stellar effective temperature of $T_{\rm
  *acc}$\footnote[2]{ We use the subscript of ``$\rm *acc$" to
  indicate that we take into account both contributions from the
  accretion powered radiation and the intrinsic radiation, summed
  together as the effective total stellar radiation.}, $\rho_{\rm
  env}$ is the envelope density, and $\Sigma_{\rm env}$ is the mass
surface density of the envelope from the dust sublimation front,
$r_{\rm sub}$, to the core radius, $R_c$.  The envelope density is
evaluated from the self-similar solution by \citet{mcl97}.  The
dust-sublimation front can be evaluated as $r_{\rm
  sub}=\sqrt{\kap_{\rm *acc}L_{\rm *acc}/[4\pi\sigma\kap_{\rm
      sub}T_{\rm sub}^4]}$, where $\kap_{\rm sub}$ is the Planck
mean opacity at the sublimation temperature of $T_{\rm
  sub}=1400\rm\:K$.  The Planck mean opacity at solar metallicity is
evaluated using the opacity table by \citet{sem03}, and it is simply
assumed to be proportional to the metallicity at lower metallicities.
At some low level of metallicity, the envelope eventually becomes optically thin
reducing the impact of radiation pressure, i.e., $f_{\rm trap}\simeq
\uptau_{\rm env}<1$.  The angular distribution of the radiation
pressure momentum is treated as spherical because only direct stellar
radiation is considered, i.e., $dp_{\rm rad}/d\Omega=p_{\rm rad}/[4\pi]$.

The momenta of the MHD disk wind and radiation pressure sweep up the envelope.
The sweeping rate $\Mdot_{\rm swp}$,
i.e., the rate for the envelope material swept up into outflow, is obtained as,
\begin{eqnarray}
\Mdot_{\rm swp} &=&-\dot{\mu}_{\rm esc}\left( M_c - M_{\rm *d} \right) \label{eq_swp},
\end{eqnarray}
where $\mu{\rm esc}=\cos\theta_{\rm esc}$.
Note that the opening angle of the outflow cavity, $\theta_{\rm esc}$, monotonically
increases with time, and thus the factor of $\dot{\mu}_{\rm esc}$ is
negative.  The mass loss by outflow sweeping is the dominant
feedback mechanism in the solar metallicity case \citepalias{KT17}.

\subsubsection{Photoevaporation}

Ionizing photons from the central massive protostar irradiate the
surface of the disk and the outflow-cavity-exposed regions of the
infall envelope creating ionized regions.  The ionized gas evaporates
from the surface if its sound speed is strong enough to escape from
the gravitational potential of the protostar.  The ionized gas is
gravitationally bounded in the inner region, while it evaporates away
from the system in the outer region.  The critical radius of this
transition is named as the gravitational radius, which is usually
evaluated as $r_{\rm g} = {Gm_{\rm *d}}/{c_{\rm H_{II}}^2}$ where
$c_{\rm H_{\rm II}}$ is the sound speed of the ionized gas
\citep{hol94,KT13}.
To take into account the repulsion by radiation pressure, we update
this formula \citep[following][]{mck08},
\begin{eqnarray}
r_{\rm g,e+d}&=&\frac{Gm_{\rm *d}}{c_{\rm H_{II}}^2} \left( 1- \Gamma_{\rm e+d} \right), \label{eq_rged} \\ 
r_{\rm g} &=& \max \left( r_{\rm sub}, r_{\rm g,e+d} \right), \label{eq_rg}
\end{eqnarray}
where $r_{\rm g,e+d}$ is the {\it tentative} gravitational radius,
$\Gamma_{\rm e+d} = {(\kappa_{\rm T}+\kappa_{\rm *acc})L_{\rm
    *acc}}/{4\pi c G m_{\rm *d}}$ is the Eddington factor considering
both electrons and dust, and $\kappa_{\rm T}$ is the opacity due to
Thomson scattering.  As a result of the dust opacity, the tentative
gravitational radius can be negative especially in higher metallicity
cases.  However, $r_{\rm g,e+d}$ should not be smaller than $r_{\rm
  sub}$ because the dust opacity is included in its estimation.
Therefore, the sublimation radius is set as the minimum value of the
gravitational radius (Eq. \ref{eq_rg}).
%\begin{eqnarray}
%r_{\rm g} &=&
%\left\{
%\begin{array}{ll}
%r_{\rm g,e} & (r_{\rm g,e} \le r_{\rm sub}),\\
%r_{\rm sub}& (r_{\rm g,e+d} \le r_{\rm sub} < r_{\rm g,e}),\\
%r_{\rm g,e+d} & (r_{\rm sub} < r_{\rm g,e+d}),
%\end{array}
%\right.\\
%r_{\rm g,e}&=&\frac{Gm_{\rm *d}}{c_{\rm H_{II}}^2} \left( 1- \Gamma_e \right),\\
%r_{\rm g,e+d}&=&\frac{Gm_{\rm *d}}{c_{\rm H_{II}}^2} \left( 1- \Gamma_e - \Gamma_{\rm d} \right),
%\end{eqnarray}
Considering the evaporation speed is the sound speed of the ionized gas,
the total photoevaporation rate can be described as,
\begin{eqnarray}
\Mdot_{\rm pe} = 2\int_{r_{\rm g}}^{r_0(M_{\rm *d})} {2\pi rX^{-1}m_{\rm H}n_0{r'}c_{\rm H_{II}}} dr', \label{eq_Mdotpt}
\end{eqnarray}
where $r_0(M_{\rm *d})$ is the collapse radius inside which the
enclosing mass was originally equal to $M_{\rm *d}$, and $n_0(r)$ is
the base density at the ionization boundary \citep{hol94}.  Based on
an accurate radiative transfer calculation, \citet{KT13} provided a
base density model in the dust-free case,
\begin{eqnarray}
n_0(r)=  c_{\rm pe} \left(\frac{S_{\rm *acc}}{4\pi \alpha_{\rm A} r^3} \right)^{1/3},
\end{eqnarray}
where $S_{\rm *acc}$ is the ionizing photon rate from the central
star, $\alpha_{\rm A}$ is the recombination coefficient for all levels
(so-called case A), and $c_{\rm pe}\simeq0.4$ is the the correction
factor used to match numerical results.  In \citetalias{KT17}, we have
extended this formula including the absorption by dust grains as
\begin{eqnarray}
n_0(r)=  c_{\rm pe} \left( \frac{S_{\rm *acc}e^{-\uptau_{\rm pe}}}{4\pi \alpha_{\rm A} r^3} \right)^{1/3},\\
\uptau_{\rm pe}(r) = \int_{r_{\rm sub}}^{r} \sigma_{\rm a,d} n_0(r') dr',
\end{eqnarray}
where $\uptau_{\rm pe}$ is the optical depth of the photoevaporation
flow due to dust grains, and $\sigma_{\rm a,d}$ is the absorption
cross section of dust grains per H nucleon.  For this cross section,
we adopt the typical value of $\sigma_{\rm a,d}=10^{-21}\rm\:cm^{2}$
from the diffuse interstellar medium for the solar metallicity case
\citep{wei01}, and reduce it by a factor of $Z/Z_\odot$ for the
lower metallicity cases.  In the evaluations of the ionizing photon
rates, $S_{\rm *acc}$, and the sound speed of ionizing gas, we take into
account of metallicity dependence by using the stellar atmosphere
model ATLAS \citep{kur91,cas04} and the spectral synthesis code CLOUDY
\citep{fer13}.

Following \citetalias{KT17}, we introduce the characteristic optical
depth of the photoevaporation flow, which is evaluated from the
physical values at the dust sublimation front:
\begin{eqnarray}
\hat\uptau_{\rm d} = \sigma_{\rm a,d} r_{\rm sub} n_0(r_{\rm sub}). \label{eq_hattau}
\end{eqnarray}
As will be seen in later sections, this characteristic optical
depth is a good measure of the strength of the photoevaporation.

\subsubsection{Net Accretion Rates}\label{sec_accretion}

As described above, several feedback processes combine to reduce the
accretion rate below the value it would take in the limit of no
feedback (Eq. \ref{eq_infall}). This reduced accretion rate is
expressed as
\begin{eqnarray}
\mdot_* = \mu_{\rm esc}\Mdot_{\rm *d} - \mdot_{\rm d} - \mdot_{\rm dw} - \Mdot_{\rm pe}, \label{eq_mdot}
\end{eqnarray}
where $\mdot_{\rm d}$ is the mass growth rate of the disk.  Since the
mass ratio of the disk to the protostar is fixed as $f_{\rm d}=1/3$
during the main accretion phase, the mass growth rate of the disk is
$\mdot_{\rm d}=\mdot_*/3$.  The first term of the right hand side
represents the infall rate, which is reduced by a factor of $\mu_{\rm
  esc}$ from its no-feedback limit (Eq. \ref{eq_infall}). Diversion of
mass into the outflow by direction injection in the disk wind and by
photoevaporation is accounted for by the final two terms.  The
evolution of the protostar and the accretion structure is solved until
accretion is cut off by the feedback processes, i.e., $\mdot_*=0$ or
the entire natal core collapses, i.e., $M_{\rm *d}=M_c$.  In the
former case, the stellar mass at its birth $m_{\rm *f}$ is set at that
moment.  In the latter case, while the main accretion phase has now
finished, accretion still continues from the remnant disk until it
dissipates.

\subsection{The Disk Dissipation Phase}

The final stage is the disk dissipation phase in which the remnant
disk accretes onto the star and/or is blown away by feedback (Figure
\ref{fig_schematics}c).  In \citetalias{KT17} this phase was ignored,
and thus the calculated final stellar masses and the SFEs were minimum
values, with maximum fractional errors of $1/3$. Now in this paper,
we extend the evolutionary calculation until the remnant disk
dissipates.

In this phase, the disk mass decreases monotonically because supply
from the core infall envelope has ended.  The rate of change of the
disk mass can be written as,
\begin{eqnarray}
\mdot_{\rm d} = -\mdot_*-\mdot_{\rm dw}-\Mdot_{\rm pe}. \label{eq_mdotd}
\end{eqnarray}
We use Equation (\ref{eq_Mdotpt})
with the maximum of the integral range is the disk radius $r_{\rm d}$
to evaluate the photoevaporation from the disk $\Mdot_{\rm pe}$.
We ignore radiation pressure in this phase since the self-shielding of the disk is expected to be efficient
\citep{tan11,kui12}.  To evaluate the accretion rate onto the star,
the $\alpha$-disk model of \citet{sha73} is introduced, which describes
the viscosity as $\nu_{\rm vis}=\alpha c_{s}H$, where $c_{s}$
is the sound speed.  Following \citet{kui10}, the viscous parameter
and the aspect ratio are fixed in the estimation of the viscosity as
$\alpha=0.3$ and $H/r=0.1$, which is equivalent to the so-called
$\beta$-viscosity model for self-gravitating disks with
$\beta\simeq0.003$ \citep{dus00}.  This assumption is reasonable while
the remnant disk has non-negligible mass compared to the stellar mass,
i.e., $m_{\rm d}/m_*\ga0.1$.  Then, the viscous accretion rate is
evaluated as $\mdot_{\rm vis}=m_{d}r_{d}/\nu_{\rm vis}$.  
Viscous accretion powers the MHD disk wind, with its mass flux still assumed to
be a fraction $f_{\rm dw}$ of the accretion onto the star, i.e., $\mdot_{\rm
  vis}=\mdot_*+\mdot_{\rm dw}=(1+f_{\rm dw})\mdot_*$.  Therefore, the
mass accretion rate onto the star is
\begin{eqnarray}
\mdot_* = \frac{\mdot_{\rm vis}}{1+f_{\rm dw}}
= \frac{m_{d}r_{d}}{(1+f_{\rm dw})\nu_{\rm vis}}. \label{eq_mdot2}
\end{eqnarray}
Note that the total mass loss rate from the star-disk system
is the sum of the MHD disk wind and the photoevaporation, i.e.,
$\mdot_{\rm *d}=-\mdot_{\rm dw}-\Mdot_{\rm pe}$.
We solve the evolutionary sequence of the star and the disk until the end of the
accretion from the remnant disk, i.e., $m_{d}=0$, and finally obtain the
stellar mass at its birth $m_{\rm *f}$.

We evaluate the SFEs from prestellar cloud cores with
$M_c=10$--$1000\msun$, $\Sigma_{\rm cl}=1\:{\rm g\:cm}^{-2}$ and
$Z=10^{-5}$--$1\zsun$.  Following \citetalias{KT17}, we define the
``instantaneous SFE'' as the ratio of the accretion rate to the
infall rate without feedback, i.e., $\varepsilon(t) \equiv
\mdot_*/\Mdot_{\rm *d}$.  The instantaneous SFE is important
especially as it can be observable in individual protostellar systems
\citep[e.g.,][]{zha16}.  However, in this series of papers, we focus
mainly on the final SFE, rather than the instantaneous SFE, to
investigate the relation between CMF to IMF.  Therefore, we
use ``SFE'' to refer to the ratio of the final stellar mass to the
initial core mass, i.e., $\sfe\equiv m_{\rm *f}/M_c$.

\begin{figure}
\includegraphics[width=85mm]{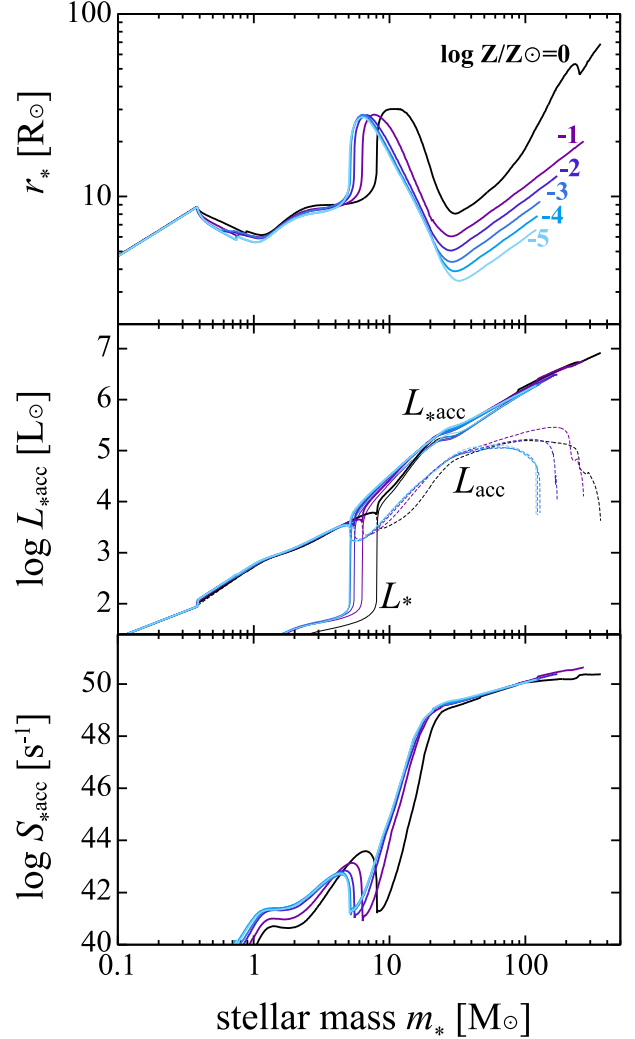}
\caption{
The evolution of protostellar radius (top), luminosity (middle) and
ionizing photon rate (bottom) at all modeled metallicities as
indicated in the top panel.
The initial core mass is $M_c=1000\:\msun$ for all cases.
In the middle panel,
the intrinsic luminosity $L_*$ (thin-solid), the accretion luminosity
$L_{\rm acc}$ (thin-dashed), and the total luminosity $L_{\rm *acc}$
(thick-solid) are shown.
}
\label{fig_prost}
\end{figure}

\section{Results}\label{sec_results}

\subsection{Accretion and Mass-Loss Histories at Various Metallicities} 

Figure \ref{fig_acc} shows the accretion histories as functions of
protostellar mass and time at various metallicities for the initial
core mass of $M_c=1000\:\msun$ embedded in a $\Sigma_{\rm cl}=1\:{\rm
  g\:cm}^{-2}$ clump environment.  The accretion rate in the no
feedback limit is also shown for reference.

At solar metallicity, the accretion rate increases as the infall rate
increases (Eq. \ref{eq_infall}).  When the stellar mass reaches around
$30\:\msun$ at a time of $5\times10^4{\rm\:yr}$, the accretion rate
starts to deviate significantly below that of the no-feedback
limit. However, the accretion rate still continues to rise until
$m_*\simeq 175\:M_\odot$ ($t\simeq1.4\times10^5{\rm\:yr}$), where the
peak accretion rate is about $2\times10^{-3}\:\msunyr$. It then
decreases as feedback becomes ever stronger at higher protostellar
masses.  The decline of accretion is mainly caused by the opening-up
of the outflow cavity as was seen in \citetalias{KT17}.  The plateau
starting at $m_*\sim250\:\msun$ ($1.8\times10^5{\rm\:yr}$) appears
because the opening angle reaches its maximum, limited by the disk
thickness.  Infall from the envelope finishes at $290\:\msun$, which
is the end of the main accretion phase
($t\simeq2.6\times10^5{\rm\:yr}$, as indicated by the vertical line in
the right panel).  In the subsequent disk dissipation phase, the
accretion rate decreases as the remnant disk dissipates and a final
stellar mass of $m_{\rm *f}\simeq359\:\msun$ is achieved by
$6.6\times10^5{\rm\:yr}$.  The SFE from the core is then
$\sfe=359\:\msun/1000\:\msun\simeq0.36$.  The final stellar mass is
higher than that obtained in \citetalias{KT17} ($290\:\msun$), because
accretion during the disk dissipation phase is newly included in this
paper.

At lower metallicities, the accretion rate and the final stellar mass
become smaller, although the other initial conditions are the same.
This is because the impact of the total feedback becomes higher at
lower metallicities, which is a trend that is opposite from that in
the classic view with idealized spherical accretion.  At
metallicities lower than $10^{-3}Z_\odot$, the accretion history is
almost identical.  In those cases, the accretion rate starts to drop
at $m_*\simeq15\:\msun$ ($4\times10^4{\rm\:yr}$) and the main
accretion phase finishes at $m_*\simeq120\msun$
($2.6\times10^5{\rm\:yr}$).  The mass accreted in the disk dissipation
phase is negligible unlike in the solar metallicity case.  The SFE at
$10^{-5}\zsun$ is $\sfe=120\:\msun/1000\:\msun\simeq0.12$ which is
lower than that at $\zsun$ by a factor of three.

\begin{figure}
\includegraphics[width=90mm]{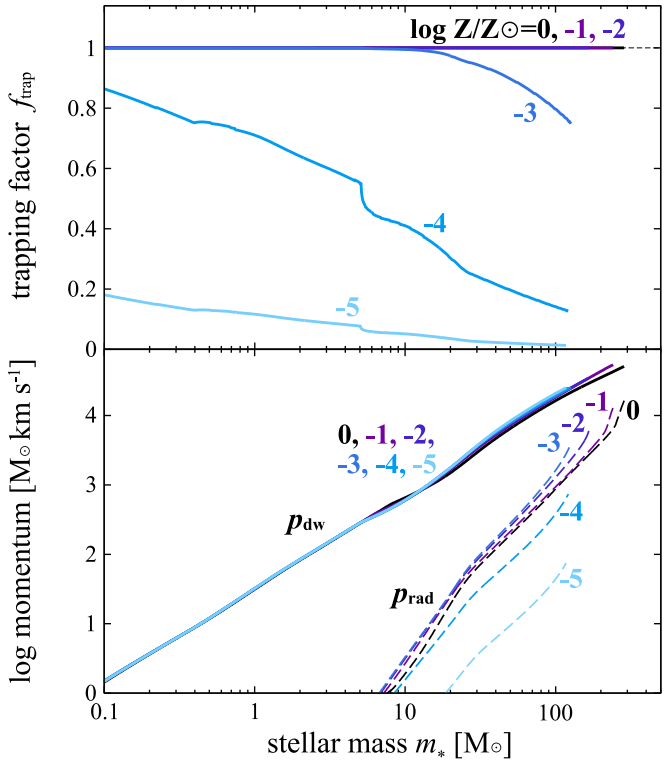}
\caption{
The evolution of the trapping factor $f_{\rm trap}$ (top) and the
momentum from the MHD disk wind $p_{\rm dw}$ and from radiation
pressure $p_{\rm rad}$ (bottom)
during the main accretion phase
at various metallicities as indicated
(for the fiducial cores with $M_c=1000\:\msun$).  At metallicities
$\le10^{-4}\:\zsun$, the trapping factor becomes lower than unity
and thus the radiation pressure momentum is weaker than in higher
metallicity cases. Note, the MHD disk wind is the main source of
momentum at all metallicities.  }
\label{fig_Prad}
\end{figure}

\begin{figure}
\includegraphics[width=90mm]{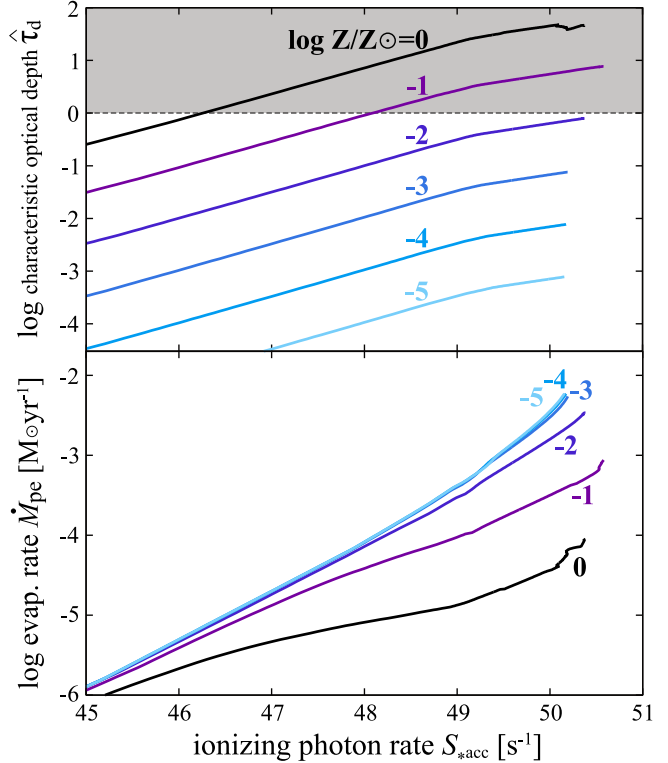}
\caption{
The evolution of the characteristic optical depth of the
photoevaporation flow $\hat\uptau_{\rm d}$ (Eq. \ref{eq_hattau})
(top) and the photoevaporation rate $\Mdot_{\rm pe}$ (bottom) at
during the main accretion phase
various metallicities as indicated
(for the fiducial cores with $M_c=1000\:\msun$).
At higher metallicities of
$\ga10^{-1}\:\zsun$, the increasing photoevaporation rate
slows down as the flow becomes optically thick $\hat\uptau_{\rm d}>1$
(shaded region in the top panel). At lower metallicities of
$\la10^{-3}\:\zsun$, the photoevaporation evolution converges to a
low metallicity limit.  }
\label{fig_PE}
\end{figure}

Figure \ref{fig_mass_loss} shows the mass-loss histories at $\zsun$
and $10^{-5}\:\zsun$ with same initial core mass of $M_c=1000\:\msun$.
As presented in \citetalias{KT17}, the dominant
mass-loss mechanism at $\zsun$ is outflow sweeping driven by the
momentum of the MHD disk wind and radiation pressure.  The sweeping
rate increases and becomes higher than the accretion rate at $30\:\msun$.
At $m_*\sim250\:\msun$, the sweeping rate drops off as the opening
angle reaches close to its maximum limit set by the disk thickness.  The
photoevaporation rate quickly increases from $m_*\sim10\:\msun$,
however, its rate of increase reduces above $20\:\msun$.  This
regulation of the photoevaporation rate is mainly caused by dust
absorption of the ionizing photons \citepalias{KT17}.

On the other hand, in the case of the low metallicity of
$10^{-5}Z_\odot$, the photoevaporation rate is not limited by dust to
be under $10^{-4}\:\msunyr$. Instead it overtakes the sweeping rate
when the stellar mass reaches $20\:\msun$.  Although the rate of
increase becomes smaller at this point, the photoevaporation mass loss
rate does still keep rising.  The sweeping rate, on the contrary,
starts to decreases earlier than is seen in the solar metallicity
case.  This is because the momentum of the disk wind is powered by
mass accretion (Eq. \ref{eq_mdotdw}), which declines due to the
efficient photoevaporation.  In this manner, the metallicity changes
which is the dominant feedback mechanism: i.e., MHD outflow sweeping
at $\zsun$ and photoevaporation at $10^{-5}\zsun$. The total impact of
feedback is thus also altered.

To reveal the causes of the metallicity dependence of the feedback
processes, we now discuss how the basic properties of protostars and
the flows driven by feedback depend on $Z$, using the results from
initial conditions of $M_c=1000\:\msun$ and covering the range
$Z=10^{-5}$--$1\:\zsun$ (the corresponding accretion histories were
shown in Figure \ref{fig_acc}).

Figure \ref{fig_prost} presents the evolution of stellar radii, $r_*$,
luminosities, $L_{\rm *acc}$, and ionizing photon rate $S_{\rm *acc}$
at various metallicities.  In the top panel, the basic evolution of
the stellar radius is same for all metallicity cases: the protostar
swells from $5$--$8\:\msun$, then returns via Kelvin-Helmholtz (KH)
contraction, and reaches the ZAMS phase at $\sim30\:\msun$.
Additionally, there are some apparent metallicity dependences.  The
swelling phase starts earlier at lower metallicity.  This is because
the swelling occurs when the opacity becomes low enough to
redistribute the interior entropy and the interior opacity is lower at
lower $Z$ \citep{hos09}.  Another difference is the radius in the
main-sequence phase is smaller at lower metallicity.  This is due to
the lower abundances of C, N and O atoms.  The KH contraction
continues until sufficient energy is produced by the CNO cycle, which
requires higher temperatures for lower CNO abundances.

\begin{figure*}
\begin{center}
\includegraphics[width=170mm]{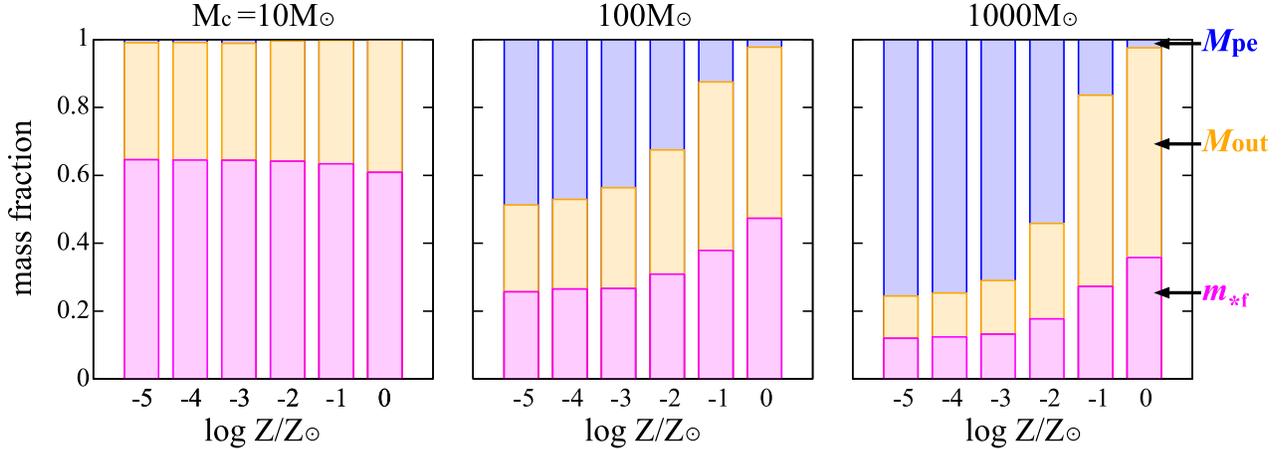}
\end{center}
\caption{
Mass fractions of final stellar mass, $m_{\rm *f}$, (pink),
outflow mass, $M_{\rm out}$, (yellow), and photoevaporated mass,
$M_{\rm pe}$, (purple), once mass accretion has ended.  Each panel shows
different initial core masses of $M_c=10,\:100,\:1000\:\msun$ (left to
right).  Each bar indicates a different metallicity of $\log
Z/\zsun=-5,\:-4,\:-3,\:-2,\:-1,\:0$ (left to right in each panel).
The metallicity dependence of the photoevaporative mass is apparent
for $M_c=1000\:\msun$, which is not seen in the $M_c=10\:\msun$ case.
}
\label{fig_mass_fraction}
\end{figure*}

Differences that are seen in the radius evolution also influence
radiation properties of $L_{\rm *acc}$ and $S_{\rm *acc}$.  In general
a smaller radius will lead to a greater accretion luminosity and a
hotter photosphere and so more H-ionizing photons per patch of the
stellar surface.  Still, as presented in Figure \ref{fig_prost} middle
and lower panels, the relative differences for different metallicities
become quite small at higher protostellar masses (although note the
larger dynamic range of these panels).  Especially, the deviation is
smaller when the stellar mass is higher than $20\:\msun$ when the
radiative feedback would be significant.  In other words, the
metallicity dependence of the protostellar evolution does not
significantly affect the strength of the radiative feedback in our
model calculations.

The top panel of Figure \ref{fig_Prad} shows the trapping factor,
$f_{\rm trap}$, (Eq. \ref{eq_frap}).  As expected, the trapping is
less efficient at lower metallicity: the trapping factor is always
$f_{\rm trap}=1$ for $Z>10^{-3}\:\zsun$, while it becomes less than
unity for $Z\le10^{-4}\:\zsun$.  As a result, the momentum feedback
due to radiation pressure is less in the lowest metallicity cases than
that in higher metallicity cases (bottom of Figure
\ref{fig_Prad}).  This trend of the trapping factor has significant
importance in the classical spherical models.  However, in our
axisymmetric model, the MHD disk wind is always the dominant source of
the momentum that drives the opening of the outflow cavity at all
metallicities (bottom of Figure \ref{fig_Prad}).  Therefore, the
metallicity dependence of the radiation pressure momentum has a minor
impact on the total feedback strength and the SFEs.

The top panel of Figure \ref{fig_PE} shows the characteristic
optical depth of the photoevaporation flow $\hat\uptau_{d}$
(Eq. \ref{eq_hattau}) and the photoevaporation rate $\Mdot_{\rm
  pe}$ as a function of the ionizing photon production rate.  In the solar
metallicity case, the characteristic optical depth reaches the
optically thick regime when $S_{\rm *acc}\ga10^{46}\rm\:s^{-1}$, and
thus the increasing rate of $\Mdot_{\rm pe}$ is suppressed.  On the
other hand, in the cases with $Z\le10^{-3}\:\zsun$, the characteristic
optical depth is always $\hat\uptau_{d}\ll1$.  As a result, the
photoevaporation rate increases smoothly and reaches higher than
$10^{-3}\:\msunyr$, which is the typical value of the accretion rate
(Figure \ref{fig_acc}).
In this way, photoevaporation has a more and more
significant impact as metallicity is lowered,
as seen in Figure \ref{fig_mass_loss}
\citepalias[see also the analytic argument in \S4.3 of ][]{KT17}.

Figure \ref{fig_mass_fraction} shows the mass fractions of the final
stellar mass, $m_{\rm *f}$, the outflow mass, $M_{\rm out}$, and the
photoevaporated mass, $M_{\rm pe}$, compared to the initial core mass at
the end of each model calculation. Results are shown for initial
conditions with $M_c=10,\:100,\:1000\:\msun$ (all for $\Sigma_{\rm
  cl}=1\:{\rm g\:cm}^{-2}$) and $\log
Z/\zsun=-5,\:-4,\:-3,\:-2,\:-1,\:0$.  The mass fraction of $m_{\rm
  *f}$ (pink bars) is equivalent to the SFE, $\sfe$.  The outflow mass
$M_{\rm out}$ is the sum of the time-integral of the sweeping rate,
$\Mdot_{\rm swp}$, and the mass-loss rate by the MHD wind, $\mdot_{\rm dw}$,
while the photoevaporated mass, $M_{\rm pe}$, is the time-integral of the photoevaporation rate, $\Mdot_{\rm pe}$.
As seen above, in the case with $M_c=1000\:\msun$ (right panel), the
outflow is the dominant feedback effect at solar metallicity, while
photoevaporation becomes significant at $Z\la10^{-2}\:\zsun$ reducing
the SFE.  Similar to Figure \ref{fig_acc}, it can be seen that all
mass fractions are similar in the low metallicity regime,
$Z\la10^{-3}\:\zsun$.  This is because photoevaporation becomes
optically thin to dust opacity at these metallicities (Figure
\ref{fig_PE}).  As the initial core mass decreases, however, the above
trend becomes weaker.  In particular, the results are almost identical
in the case of $M_c=10\:\msun$: the outflow is the dominant feedback
mechanism and photoevaporation is negligible.  In this lower mass
case, the stellar mass is too low to have significant radiative
feedback, and thus the effective feedback is only from the MHD disk wind,
which is assumed to be independent of the metallicity.  Note that the
SFE at lower masses are higher than that in \citetalias{KT17}, because
the mass accreted during the disk dissipation phase is now included.

\subsection{Star Formation Efficiencies at Various Metallicities}\label{sec_Zdep}

We have seen that the impact of feedback depends on
metallicity and also on initial core mass.  To show these trends
more clearly, the SFEs at various metallicities are plotted as a
function of the final achieved stellar mass $m_{\rm *f}$ in Figure
\ref{fig_sfe}.

First, the SFE decreases with the final stellar mass at all
metallicities, because radiative feedback becomes stronger in
higher-mass star formation.  This trend is true even in the solar
metallicity case in which the MHD disk wind is the dominant feedback
rather than radiative feedback \citepalias{KT17}.  Second, the SFE for
$m_{\rm *f}\la10\:\msun$ is nearly independent of metallicity, since
only MHD disk winds are effective feedback in this low-mass regime
(left panel of Figure \ref{fig_mass_fraction}).  Finally and most
importantly in this paper, the SFE for $m_{\rm *f}\ga20\:\msun$ is
lower for the lower metallicity cases, and approaches converged
results that are independent of metallicty once $Z\la10^{-3}\:\zsun$.
This is mainly caused by the metallicity dependence of the
photoevaporation rate at higher metallicities, where it becomes
suppressed by the presence of dust (\S\ref{sec_Zdep}).  This trend of
lower SFE at lower metallicity may have potential important
implications for systematic variation of the high-mass end of the IMF
with galactic environment.

The obtained SFE can be fitted by
\begin{eqnarray}
\sfe \simeq 0.60 \left( \frac{M_c}{12\:\msun}  \right)^{\varepsilon'},\\
\varepsilon' \simeq -0.11+ 0.084 \max\left( \log \frac{Z}{\zsun}, -3 \right), \label{eq_epp}
\end{eqnarray}
which agrees with our numerical results within a maximum error of
$(\Delta \sfe)_{\rm max}=0.03$ over the wide ranges of parameters of
$M_c=10$--$1000\:\msun$ (all for $\Sigma_{\rm cl}=1\:{\rm
  g\:cm}^{-2}$) and $Z=10^{-5}$--$1\zsun$.  This simple fitting
formula (and potential generalizations for different $\Sigma_{\rm
  cl}$) can be applied as a sub-grid model to large-scale simulations
of star formation that resolve formation of massive prestellar cores.

We note one more interesting finding from our model calculations.
Although the SFE decreases with stellar mass at all metallicities, its
decline does not show any abrupt cut-off up to about $300\:\msun$.  In
other words, for our adopted initial conditions, there is no evidence
for an upper limit to the birth mass of stars being caused by
feedback, in contrast to spherical models \citep{hos09b,fuk18}.

\section{Discussion}\label{sec_discussion}

\subsection{Implications for IMF Variations}\label{sec_imf}

Massive stars are short-lived and thus constitute the main source of
heavy elements injected into the interstellar and intergalactic
media, especially in the early universe.  Additionally, their
feedback by strong radiation and supernovae affects the dynamical and
chemical evolution of galaxies.  Therefore, the stellar IMF must be
known to predict element production, the impact of feedback, and
the formation rate of black holes.  However, the universality of the
IMF is still under investigation, and it is uncertain if it
depends on environment and metallicity \citep[e.g.,][]{bas10}.

We thus discuss the importance of feedback processes and
their metallicity dependence to IMF variation based on our model calculations.
Our model shows that SFE from a core is lower when
forming stars of higher mass and under lower metallicity conditions
(\S\ref{sec_Zdep}).  Considering the IMF to be the multiplicative
product of the combination of CMF and SFE, the shape of the high-mass
end of the IMF is then expected to deviate from the CMF shape in
contrast to present-day low-mass star formation
\citep{and10,kon10,che18}.

We can quantitatively link the IMF and CMF based on the obtained
SFEs \citep{nak95,mat00}.  Assuming the CMF and SFE are power-law
distributions of $d\N / d\ln M_c = \N_0 (M_c/M_0)^{-\alpha_c}$ and
$\sfe=\varepsilon_0(M_c/M_0)^{\varepsilon'}$, where variables with a
subscript of ``$0$'' indicate the normalized values, the IMF can be
written as,
\begin{eqnarray}
\frac{d\N}{d\ln m_{\rm *f}} = \frac{\varepsilon_0^{\alpha_c/(1 +\varepsilon')}\N_0}{1+\varepsilon'} \left(  \frac{m_{\rm *f}}{M_0}\right)^{-\alpha/(1+\varepsilon')}. \label{eq_imf}
\end{eqnarray}
As the exponent of $\varepsilon'$ is a negative value of
$\sim-0.11$--$-0.36$ (Equation \ref{eq_epp}) the upper-end IMF slope is steeper
than the CMF slope by a factor of $\sim1.1$--$1.6$ depending on the
metallicity.  Thus, the number of massive stars is lower than the
simple estimation with a constant SFE.

As an example, assuming a CMF slope of $\alpha_c=1.35$, similar in
shape to the \citet{sal55} IMF from $\sim1$ to $\sim10\:M_\odot$, we
evaluate the upper-end IMF at various metallicities based on the
fitting of our SFEs (Figure \ref{fig_imf}).  Here the initial CMF is
normalized at $10\:\msun$, i.e., $d\N / d\ln M_c =
(M_c/10\:\msun)^{-1.35}$.  In the solar metallicity case, the upper-end IMF
slope is then $1.53$, which is a little steeper than the assumed CMF slope.
The IMF slope becomes steeper as metallicity decreases, and it
converges to $\alpha_c/(1-0.36)\sim2.1$ at around $10^{-3}\:\zsun$
as a result of the metallicity dependence of feedback processes
(\S\ref{sec_Zdep}).  Due to this difference of the IMF slope, the
number of stars with $30$--$100\:\msun$ at $10^{-5}\:\zsun$ is $2.2$
times smaller than the number at $\zsun$, and that factor is $4.6$ for
the mass range of $100$--$300\:\msun$ (assuming the same initial CMF
is applied).  In this manner, our model predicts that massive stars
are relatively harder to form at lower metallicity, especially
$\la10^{-3}\:\zsun$.

\begin{figure}[t]
\begin{center}
\includegraphics[width=85mm]{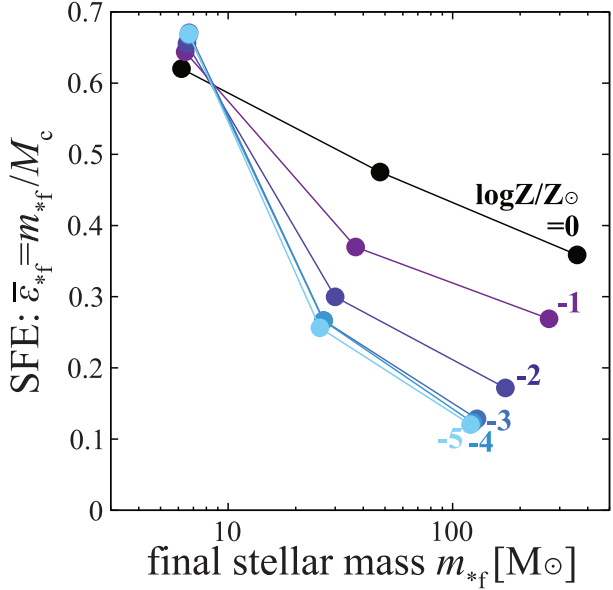}
\end{center}
\caption{
SFEs at various metallicities are plotted as a function of the
stellar mass at their birth $m_{\rm *f}$.  The SFE is lower at higher
masses and at lower metallicity due to stronger feedback.}
\label{fig_sfe}
\end{figure}

\begin{figure}[t]
\begin{center}
\includegraphics[width=85mm]{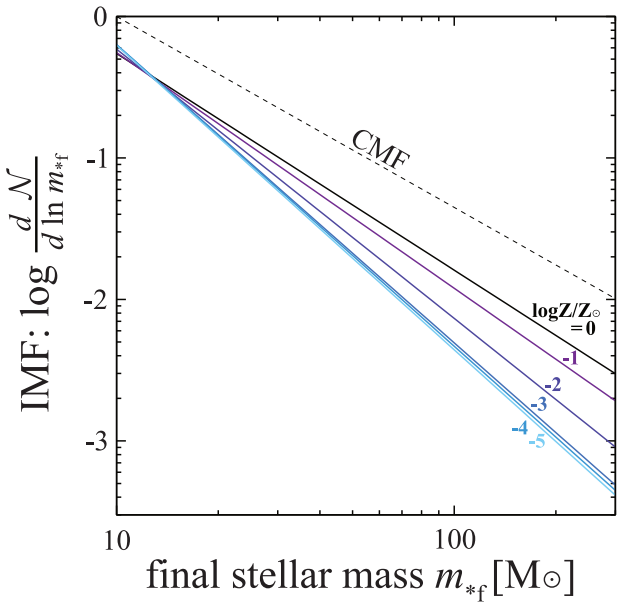}
\end{center}
\caption{
The evaluated IMFs at various metallicities based on our SFE model
under the assumption that the high-mass end of the CMF has a power law
index equal to the Salpeter value of $\alpha_c=1.35$ (dashed line).
The upper-end IMF is steeper at lower metallicities.
}
\label{fig_imf}
\end{figure}

An interesting observed feature of the high-mass end of the IMF is the
maximum stellar mass in young massive clusters.  \citet{fig05}
suggested the upper-mass limit of $150\:\msun$ based on the Arches
cluster near the Galactic Center.  However, recent observations of the
30 Doradus star-forming region in the Large Magellanic Cloud (LMC)
reported very massive stars whose initial masses are estimated to be
as high as $200$--$300\:\msun$ \citep{cro10,cro16}.  \citet{sch18}
reported that the IMF in 30 Doradus has an excess of massive stars
with the slope of $0.90^{+0.37}_{-0.26}$, which is shallower than the
Salpeter value of $1.35$.  Since the metallicity in the LMC is about
$0.4\:\zsun$, one might suppose this difference of the maximum stellar
mass may come from the effect of the metallicity dependence of the
feedback.  However, our model showed that the impact of total feedback
is stronger at lower metallicity leading to the opposite trend.  To
reconcile with model results, we therefore speculate that the initial
CMFs in these regions may have been different, i.e., there may have
been a CMF cut-off in the Arches, and the CMF slope in 30 Doradus may
have been shallower than the observed IMF.  Our model predicts the CMF
slope of the 30 Doradus is as shallow as $\sim0.77^{+0.32}_{-0.22}$
assuming a metallicity of $0.4\:\zsun$ (Equations \ref{eq_epp} and
\ref{eq_imf}).  This speculation suggests that the CMF depends on
environmental properties \citep[e.g.,][]{che18,mot18,liu18},
or that other mechanisms,
like a small number of protostellar mergers, might influence the formation of the
most massive stars.

While the CMF is not the main question in this study, the metallicity
dependence of the CMF is crucial to fully understand potential IMF
variation in different galactic environments and thus over cosmic
history.  In primordial star formation, the typical core mass is as
high as $1000\:\msun$ or more due to the lack of an efficient coolant
\citep[e.g.,][]{bro04}.  The analytic model calculation by
\citet{omu05} showed that, at the metallicity of $\ga10^{-5}\:\zsun$,
cloud fragmentation induced by dust cooling mainly creates low-mass
fragments with $\la1\:\msun$, rather than massive ones \citep[see also
  simulations by][]{dop13}.  Still, a full understanding of the CMF
likely requires accounting for nonthermal processes, such as
turbulence and/or magnetic fields \citep[e.g,][]{pad17,li18}.

Another important process to determine the stellar birth mass is disk
fragmentation.  Although massive cores have many thermal Jeans masses
at solar metallicity, catastrophic fragmentation is suppressed by
radiative heating by high accretion luminosity and efficient
angular-momentum transport by magnetic breaking
\citep[e.g.,][]{kru07,com11}.  However, small amounts of fragmentation
may still occur forming binaries/multiple systems.  Indeed,
\citet{san12} showed that the more than $70\%$ of observed massive
stars have close companions that eventually exchange mass.
\citet{KT14} analytically studied the metallicity dependence of the
self-gravitational instability of protostellar disks.  They found that
the protostellar disk is strongly unstable due to efficient dust
cooling at $10^{-5}$--$10^{-3}\:\zsun$ with typical accretion rates
of massive star formation, i.e., $10^{-4}$--$10^{-3}\:\msunyr$.
However, this analysis did not allow for the effects of magnetic
fields on disk fragmentation.

\subsection{Caveats}

We have adopted a semi-analytic model that is still highly simplified and
idealized, even though it already has some agreements with
observations \citep{zha13a,KT16,deb17}.  Below, we discuss some
caveats of our modeling.

As described above, we considered only single star formation not
allowing for fragmentation.  In the formation of present-day massive
stars \citep{kru09,ros16} and primordial stars \citep{sta10,sus14},
three-dimensional simulations suggest that fragmentation of
protostellar disks leads to the formation of multiple systems.
However, magnetic fields are expected to suppress fragmentation
\citep{mac08,com11}, and so our model may apply in this limit.  We
expect that our model is still quantitatively appropriate as long as
the total stellar mass is dominated by that of the most massive star.
On the other hand, if the system contains two or more similar mass
stars, our model would need modifications.  The momentum rate from MHD
disk winds is roughly proportional to the total accretion rate
(Equation \ref{eq_pdotdw}), and thus the number of stars would not
significantly alter the MHD wind feedback.  In contrast, the total
radiative feedback would become weaker in multiple systems, because
the luminosity and the ionizing photon rate increase nonlinearly with
the stellar mass at least for $\la100\:\msun$.  Therefore, the total
feedback could be somewhat weaker.

We adopted the same dust model as at solar neighborhood even for lower
metallicity cases, while the dust properties are thought to be
different in the early universe. For example, dust grains in the early
universe are considered to be produced in supernovae and affected by
reverse shocks.  These are thought to reduce the fraction of metals
in the dust phase destroying especially smaller-size grains
\citep{noz07}.
The metal fraction in dust would increase to
the solar-neighborhood value during the prestellar collapse phase
\citep{chi13}, which may justify our assumption in this work.  However,
if the dust distribution tends to be biased to large sizes, then the
opacity for the ionizing photon would be smaller than our assumed value.
Thus, the metallicity at which the photoevaporation becomes
significant could be somewhat higher than $10^{-2}\:\zsun$ compared to our
model result. However, the feedback impact at $\le10^{-3}\:\zsun$ of our model
would not enhanced by this fact, because it already reaches the saturation level of the low metallicity limit.

Finally, we note that observational tests are needed to confirm the
reliability of complex theoretical models.  We have applied the previous
versions of our models to make predictions on observations of massive
protostars at infrared \citep{zha11,zha13,zha14,zha18} and at radio
\citep{KT16,KT17a}, and also compared them to observations \citep{zha13a,deb17}.
We will perform the radiative transfer
predictions of the feedback models that we have presented here, and
test (at least near solar metallicity cases) with current and future observations,
including with {\it Stratospheric Observatory For Infrared Astronomy (SOFIA)},
{\it Very Large Array (VLA)} and {\it Atacama Large Millimeter/ submillimeter Array (ALMA)}.

\section{Conclusions}\label{sec_conclusions}

Massive stars are thought to have been astrophysically important since
the times of the first stars. Thus we have investigated the impact of
several feedback mechanisms in massive star formation and evaluated,
by semi-analytic methods, the star-formation-efficiencies (SFEs) from
prestellar cloud cores.  Previously we focused on the formation of
present-day massive stars at solar metallicity in \citet{KT17} (Paper
I). Here we have extended the model to cases with various
metallicities of $Z=10^{-5}$--$1\:\zsun$, as one measure of the effects
of galactic environment and cosmic evolution.

We found that the total impact of feedback and which process dominates
depends on metallicity.  Radiation pressure, which has been
regarded as the crucial barrier for present-day massive star formation,
has a relatively minor impact over all the metallicity range.  At solar
metallicities, the MHD disk wind is the dominant mechanism providing a
major portion ($\ga90\%$) of the outflow momentum.  As the
metallicity decreases, photoevaporation becomes stronger and reduces
the SFE, because dust attenuation of ionizing photons is
inefficient.  This metallicity dependence saturates at around
$10^{-3}\:\zsun$.

The obtained SFE from a given core decreases in the formation of
higher-mass stars at all metallicities because their feedback is stronger.
Moreover, this SFE decline is steeper at lower
metallicities due to more efficient photoevaporation (Figure \ref{fig_sfe}).
If the initial CMF is described with the Salpeter index of $1.35$,
our model predicts that the number fraction of stars with
$30$--$100\:\msun$ ($100$--$300\:\msun$) at $10^{-5}\:\zsun$ would be
$2.2$ ($4.6$) times smaller than that at $\zsun$.  We note that our
modeling does not show any clear truncation of SFE at the highest masses.
This means that the upper mass limit of stars (if it exists) is not
determined by feedback processes and that this applies for all the
metallicities we have explored.

%%%%%%%%%%%%%%%%%%
\section*{Acknowledgments}
The authors thank Taishi Nakamoto, Kengo Tomida, and Kazunari Iwasaki
for fruitful discussions and comments.  This work was supported by
NAOJ ALMA Scientific Research Grant Numbers 2017-05A.
J.C.T. acknowledges support from NSF grants AST 1212089 and AST
1411527.
Y.Z. acknowledges support from RIKEN Special Postdoctoral Researcher Program.

%\appendix

\clearpage

\end{document}